\documentclass{mem}
\usepackage{natbib}
\usepackage{newtxtext,newtxmath}
\usepackage{balance}
\usepackage{graphicx}
\usepackage{flushend}
\usepackage{amsmath} 
\usepackage[breaklinks]{hyperref}
\idline{000}{000}

\usepackage{microtype}
\usepackage{booktabs}
\usepackage{threeparttable}
\usepackage{enumerate}
\usepackage{multicol}
\usepackage{multirow}

\usepackage[dvipsnames]{xcolor}

\newcommand{\jwst}{\textit{JWST}}

\begin{document}
\def\teff{$T\rm_{eff }$}
\def\kms{$\mathrm {km s}^{-1}$}
\def\Msun{$M_{\odot}$}

\title{GALACTIC CENTERS AS TRACERS OF GALAXY EVOLUTION}
\subtitle{Zooming from galaxy large-scale properties into their central massive black holes}

\author{
F.~Pinna\inst{1,2} 
\and N.~Hoyer$^{3}$
\and S.~Thater
\and A. Mastrobuono Battisti
\and A. Feldmeier-Krause\inst{4}
\and E. Arjona-Gálvez\inst{1,2}
\and A. Askar\inst{8}
\and R.~C.~Levy\inst{9}
\and A. de Lorenzo-Cáceres\inst{2,1} 
\and I. Márquez\inst{10}
\and K.~Fahrion\inst{4}
\and I.~Martín-Navarro\inst{1,2}
\and C. de Sá-Freitas\inst{11}
\and R. Schödel\inst{10}
\and S. Bonoli\inst{12,13}
\and A. Chaturvedi \inst{14}
\and N. Ryde \inst{15}
}

\institute{
Instituto de Astrof\'isica de Canarias, calle V\'ia L\'actea s/n, E-38205 La Laguna, Tenerife, Spain, \email{francesca.pinna@iac.es}
\and
 Departamento de Astrof\'isica, Universidad de La Laguna, Avenida Astrof\'isico Francisco S\'anchez
s/n, E-38206 La Laguna, Spain\\
{\it The remaining affiliations can be found at the end of the paper.}
}

\authorrunning{Pinna et al.}

\titlerunning{Galaxy centers}

\date{Received: XX-XX-XXXX (Day-Month-Year); Accepted: XX-XX-XXXX (Day-Month-Year)}

\abstract{
During the workshop ''Galactic Centers as tracers of galaxy evolution'' (Lorentz Center, Leiden), we reviewed one of the most debated questions in the field of galaxy evolution: the connection between galactic central regions and the evolution of galaxies at a global scale. 
We gathered different research communities 
to reveal the connection channels between central massive black holes (MBHs), nuclear structures, 
and evolutionary processes at a global scale, such as gas inflows and outflows and galaxy-galaxy interactions. Special focus was dedicated to MBH seeding, formation and growth, active galactic nuclei feedback and the formation of nuclear stellar structures, 
without missing a global perspective. Participants from these different fields introduced us to the state of the art in numerical simulations, as well as cutting-edge observational techniques. 
In addition to achieving a deep understanding of these topics, we set up plans to exploit 
future facilities such as the Extremely Large Telescope (ELT). 
In this paper, we briefly summarize the highlights of the workshop, identify the open questions that emerged during the workshop discussions, and discuss what observations and simulations will be needed to address them. 
}
\maketitle{}

\newpage
\section{Introduction} \label{sec:intro}

In the last few decades, the astrophysical community has made great progress in understanding the formation and evolution of galaxies, shedding light on the combination of internal and external processes, such as secular evolution \citep[e.g.,][]{Kormendy2013, Kormendy2015} and galaxy mergers \citep[e.g.,][]{Lotz2008, Hopkins2010, Chilingarian2010}, which have resulted in the present day diversity of morphologies (e.g., \citealt{Pillepich2018}). 

Complexity is also prominent in the central regions of galaxies, where a variety of distinct stellar systems form, often coexisting with central massive black holes (MBHs). Together with the gas in the central regions, these stellar systems are important tracers of the dynamical state and overall evolution of the host galaxy. Studying these structures not only reveals their formation and growth but can also point to the mechanisms that funnel stars and gas to the center, which can lead to the seeding and growth of MBHs and to the appearance of active galactic nuclei (AGNs, e.g., \citealt{Bonoli2014, Neumayer2020}). 
When MBHs undergo an active phase, they interact with the host galaxy through black-hole feedback. This can be positive and foster or trigger star formation, or negative and be responsible for quenching \citep{Combes2021, Cazzoli2017, RamosAlmeida2023}. The tight connection of galactic nuclei with large-scale processes makes them key to understanding galaxy evolution in general. 

Nuclear star clusters (NSCs) and nuclear stellar disks and rings (NDRs), are the most prominent stellar structures in galactic nuclei. The latter may also be observed in the forms of nuclear bars or spirals \citep{Marquez2000, Martini2001, deLorenzoCaceres2020, Nitschai2020}. NSCs are compact, with sizes of a few parsecs, and come in a variety of shapes (e.g., \citealt{SanchezJanssen2019a, Neumayer2020}), while NDRs can be more extended, up to a few hundred parsecs, and are rotationally supported structures \citep{Pizzella2002, Gadotti2020}. With small sizes and masses between $10^{5}$ and $10^{8}$\Msun, NSCs and NSDs are very dense. Between 60 and 80\% of galaxies with stellar masses between $10^{8}$ and $10^{10}$ \Msun\, host an NSC, while NDRs have been found in about 20\% of disk galaxies \citep{Comeron2010, Neumayer2020}. 
The distinction between these structures is sometimes difficult, and in fact they have much in common with each other: sometimes similar masses, ages and metallicities, and complex stellar populations.

NSCs can form mainly via two different channels: 1) the inspiral of star clusters, formed elsewhere, towards the center due to dynamical friction and 2) in-situ star formation after gas inflow. These scenarios are not mutually exclusive, and several observational studies have suggested that both mechanisms contribute to the formation and growth of NSCs (e.g. \citealt{Fahrion2021a, Fahrion2022a, Fahrion2024}). However, there is no overall agreement on their relative contribution, nor on the dominant formation scenarios for galaxies with different morphological types \citep{Pinna2021}. Most studies on NDRs support an in-situ formation from gas funneled to the center by a bar, although an external origin has been invoked for those showing counter-rotation \citep{Bittner2020a}. 

The coevolution of galaxies and their nuclei is also traced by peculiar objects that are thought to be the stripped remnants of their progenitor galaxies, as a consequence of past galaxy-galaxy interactions. The most massive globular clusters in the Milky Way (e.g., M54 and Omega Centauri) show complex kinematics and stellar populations \citep{AlfaroCuello2019a,Clontz2024} and likely host MBHs in their centers \citep{Haeberle2024b}. Their properties, similar to NSCs, suggest they were the stellar nuclei of accreted dwarf galaxies. On the other hand, ultra-compact dwarf galaxies (UCDs) are compact stellar systems more massive than the most massive globular clusters, and more compact than dwarf galaxies of similar masses. They also show complex stellar populations and often an MBH in their center, which is over-massive (as predicted by the relation between MBH mass and galaxy mass, e.g.,\citealt{FerrareseMerrit2000}). This can be explained if they are the nuclear remnants of much more massive progenitors that were tidally stripped \citep{Seth2014, Ahn2018, FerreMateu2021}.

As mentioned above, the evolution of nuclear stellar and gaseous structures, especially NSCs, appears to be intimately connected to MBHs. In particular, both NSCs and central MBHs share mass scaling relations with their host galaxy (e.g., \citealt{Ferrarese2006}), suggesting that the formation of both is driven by processes that affect the evolution of their entire host galaxy. The very high density of NSCs favors merging events of objects such as lower-mass BHs, producing gravitational waves. These events may give birth to potential seeds of MBHs (intermediate mass black holes, IMBHs), which later grow through gas accretion and tidal disruption events, favored by the dense NSC environment (e.g., \citealt{Neumayer2020}). Thus, MBHs may form and grow in NSCs, while the same large-scale processes (e.g., rejuvenation after galaxy mergers) may play a key role in both triggering nuclear starbursts and feeding MBHs \citep{Levy2022, Levy2024, Askar2021, Askar2023, Bellocchi2023}. Gas inflow and subsequent in-situ star formation is one key mechanism to form nuclear structures, while AGN winds and outflows can wipe out all gas at galactic centers and prevent in-situ star formation. Thus, the fueling cycle of MBHs is intimately connected to the formation of stellar structures in the center of galaxies. 

Many unanswered questions remain on these topics, which have been traditionally studied separately. 
Examples of these questions are: 
\begin{itemize}
    \item What do different nuclear stellar structures (at different scales) have in common and what drives their diversity? 
    \item What is the link between the formation history of nuclear structures and galaxy properties? What is the role of external processes or AGNs in the formation of nuclear structures? 
    \item Why do some galaxies form a NSC and not a MBH and vice versa? When they coexist, which one formed first?
\end{itemize}

We organized a workshop at the Lorentz Center, ''Galactic Centers as tracers of galaxy evolution'', with the purpose of gathering scientists with expertise in all these different topics, and shed more light on these unanswered questions. 

\textbf{Outlook.} In this workshop, we brought together both theorists and observers to shed light on the connection between the key mechanisms that shape galaxy centers. In particular, typical processes of galaxy evolution at large scale, feeding and feedback cycle of MBHs, and the formation of stellar structures at the center of galaxies. On the observational side, integral-field spectroscopy coupled with adaptive optics has recently allowed high-quality studies of the links between galactic nuclei, MBHs and host galaxy evolution (e.g. \citealt{Lyubenova2013, Lyubenova2019, Pinna2021, Thater2023}). 

Nevertheless, at the moment, our capability to spatially resolve galactic centers is still limited to relatively nearby galaxies. Meanwhile, high-resolution observations with the James Webb Space Telescope (JWST) are providing an unprecedented view of stellar structures at a small scale (of few parsecs, \citealt{Hoyer2023, Fahrion2024}) and allowing to expand toward the origin of the universe, exploring the properties of the first galaxies (e.g., \citealt{Uebler2023, Uebler2024}). In addition, gravitational waves have opened a new window on galaxy evolution at small scales, and are expected to revolutionize astrophysics in the next decades \citep{LISA2023}. In particular, in the next decades, they will allow us to study in detail the merger of MBHs, which are intimately connected to galaxy evolution. From the theoretical perspective, cutting-edge numerical hydrodynamic simulations (cosmological or with idealized initial conditions) can now zoom into smaller scales and resolve gas inflows and outflows in unprecedented detail (e.g., \citealt{Hopkins2018}), while new generations of cosmological simulations are planned. 

Recent observational progress and theoretical advances, which were extensively discussed during our workshop, are paving the way for the advent of the Extremely Large Telescope (ELT). With its spectacular aperture of almost 40 meters and coupled to superb integral-field spectrographs such as HARMONI, it will be ideal to zoom into the centers of galaxies located in a wide range of distances from us (e.g., \citealt{Gullieuszik2014, Nguyen2025}), allowing accurate spatially resolved measurements of nuclear gas and stellar properties at an unprecedented spatial resolution. 
Our workshop was the right environment to plan the new challenges for the expected ELT revolution, and the discussions helped us define clear directions for future observing and funding proposals.

\textbf{Key points.} We summarize here the key points that were included in the scientific program. 
\begin{itemize}
\item Stellar nuclei and AGN co-evolution with their host galaxies: positive or negative impact of the MBH feedback cycle on star formation in the central few hundred parsecs and on the evolution of nuclear stellar structures.
\item
Evolutionary connection of nuclear structures such as NSCs and NDRs with larger-scale stellar and gaseous structures in galaxies.
\item
The role of external processes (such as galaxy-galaxy interactions) in favoring star-cluster accretion and migration to the galactic center, and gas accretion triggering nuclear star formation.
\item
Dominant formation channels for NSCs in galaxies of different masses, morphological types and star-formation rates.
\item
Formation scenarios for NDRs and their coexistence with MBHs and NSCs.
\item
Insights from the detailed study of the Milky Way center. 
\item
Properties of stripped nuclei (very massive globular clusters, UCDs), coexistence with MBHs and what they teach us about their progenitor galaxies. 
\item
Mergers of lower-mass black holes (BHs) in galactic centers, as a channel to form MBHs, and their detection through gravitational waves.
\item
Growing MBHs through tidal disruption events.
\item
How new techniques 
and instruments are revolutionizing this field. 
\end{itemize}

\section{Nuclear structures}
\label{sec:nuclear_structures}

\begin{figure*}
\centering
\includegraphics[width=0.7\textwidth]{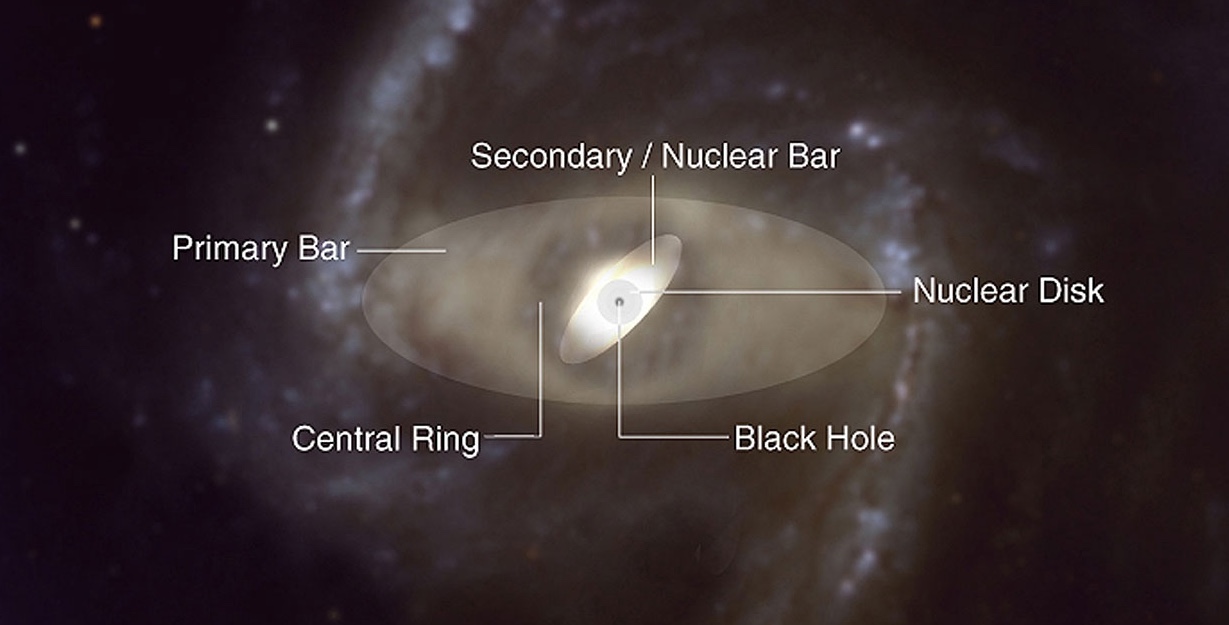}
\caption{Schematic overview of the various components of a double-barred galaxy. Credit of the Figure is by ESO.}  
\label{fig:nscformation}
\end{figure*}

We present in this section an overview of our current understanding of the properties of stellar structures of slightly different spatial scales in the nuclear regions of galaxies, emerging from the results presented during our workshop. 
Table~\ref{tab:properties} contains a brief summary of the most basic parameters of each type of nuclear component. 
Further subsections detail how to determine these values, how each nuclear structure may form, and how they may relate to each other, zooming from the largest structures into the smallest ones. 

\begin{table*}
  \scriptsize
  \caption{%
    Properties of nuclear structures.
  }
  \begin{center}
    \begin{threeparttable}
      \begin{tabular}{lllll}
        \toprule
        \multicolumn{1}{c}{\multirow{ 2}{*}{Name}} & \multicolumn{1}{c}{Typical mass} & \multicolumn{1}{c}{Typical size} 
        & \multicolumn{1}{c}{\multirow{ 2}{*}{Formation requirement}} & \multirow{ 2}{*}{References}\\
        \cmidrule(lr){2-2}
        \cmidrule(lr){3-3}
         & \multicolumn{1}{c}{$[\mathrm{M}_{\odot}]$} & \multicolumn{1}{c}{$[\mathrm{pc}]$} 
         \\
        \midrule
        \multirow{ 2}{*}{Central black hole} & \multirow{ 2}{*}{$10^{4}$-$10^{10}$} & \multirow{ 2}{*}{$0.1$-$300$\tnote{(a)}} & \multirow{ 2}{*}{Accretion / seed} & \citet{Kormendy2013}\\
         & & & & \citet{Gebhardt2011} \\
        \multirow{ 2}{*}{Nuclear star cluster} & \multirow{ 2}{*}{{$10^{4}$-$10^{8}$}} & \multirow{ 2}{*}{{$1$-$50$}} 
        & {Accretion (of star clusters or gas)} & \citet{Neumayer2020}\\ 
         & & & {high nuclear gas density} & \citet{Georgiev2014}\\
        \multirow{ 2}{*}{Circumnuclear ring} & \multirow{ 2}{*}{} & \multirow{ 2}{*}{$200$-$1000$} & \multirow{ 2}{*}{} & \citet{Comeron2010}\\ 
         & & & & \citet{Busch2017}\\
        {Inner bar} & {$10^{7}$-$10^{9}$} & {$100$-$300$}  & {Presence of a large-scale disk} & \citet{Gadotti2020}\\
        {Nuclear stellar disk} & {$10^{8}$-$10^{10}$} & {$100$-$300$}  & {In general, presence of a bar} & \citet{Schultheis2025}\\
        \bottomrule
      \end{tabular}
      \begin{tablenotes}
        \item[(a)] Sphere of influence. Depends on the presence and properties, thus the velocity dispersion, of other nuclear components as $r_{\mathrm{SOI}} \propto M_{\mathrm{BH}} \sigma^{-2}$.
      \end{tablenotes}
    \end{threeparttable}
  \end{center}
  \label{tab:properties}
\end{table*}

\subsection{Nuclear stellar disks}
\label{subsec:nuclear_stellar_disks}

Nuclear discs follow exponential profiles and have sizes between 100 and 1000~pc, (which correlate with the bar length \citealt{deSaFreitas2025a}) and masses between $10^8$ and $10^{10}$~\Msun \citep{Gadotti2025a}. 
Nuclear stellar rings are believed to be just a phase of the inside-out formation of nuclear disks (e.g., \citealt{Bittner2020a}, see also below). 

NDRs emerge as one of the consequences of the presence of bars on the evolution of the (disk) galaxy. They usually need a bar structure to form (although several nuclear disks were detected in unbarred galaxies, \citealt{Schultheis2025}, and they are located at the inner Lindblad resonance (see, for example, the simulation by \citealt{Sormani2024a}). Bars induce tangential forces and shocks, causing gas inflow, which finally end up forming stars in the inner region. The new bar-built stellar structures are expected to be rotationally supported \citep[e.g.,][]{Shlosman1989,Marquez2003,Emsellem1996} and can tell us when bars formed \citep{Gadotti2015}. 
The presence of a disk is needed to form a bar, implying that the disk must have formed before the bar. If the bar forms from stars previously belonging to a thin disk, it is believed that no major mergers can have happened after that. While it is unclear how long it takes between the last major merger and the formation of the main disk of a galaxy, after disk formation, the bar can be formed quickly, and just after that, a nuclear disk. Therefore, the first generation of stars in the nuclear disk would be approximately as old as the bar. 
Thus, while the time of dynamical bar formation cannot be extracted from their stellar populations (most of bar stars belonged to the disk when they formed), the age of the youngest stars in NDRs can be used to date bar formation. Bar formation would be dated when the star formation history of the nuclear disk shows that new stars were formed and start to dominate over the existing ones (extracted in an annulus, \citealt{deSaFreitas2023a}). 

From the TIMER survey \citep{Gadotti2019a}, targeting 24 nearby, Milky Way-mass barred galaxies, we learned that nuclear disks and rings are rapidly rotating structures, with low velocity dispersion, and are more rotationally supported than the main disk of the galaxy at the same radii, and are aligned to it. They show an anti-correlation between velocity and the third moment $h_3$ of the line-of-sight velocity distribution (LOSVD), a signature of near-circular orbits, and high values of $ h_4$, suggesting that they are a component distinct from the main underlying disk \citep{Gadotti2020}. 
Nuclear disks show relatively young, metal-rich stellar populations compared to their surrounding galaxy, and lower values of [$\alpha$/Fe], supporting their bar-driven origin \citep{Bittner2020a}. 
Well-defined radial gradients were uncovered in the nuclear disks by the TIMER survey, with decreasing stellar ages and metallicities, increasing [$\alpha$/Fe] abundances. A single slope from the centre to their outer edges suggests that nuclear disks are continuous
components all the way to the centre, which were formed in a continuous way, and these gradients suggest an inside-out formation scenario \citep{Bittner2020a, deSaFreitas2023a}. These findings support a common origin of NDRs, with nuclear rings as part of the continuous formation of a nuclear disk. Inside-out formation of NDRs was also witnessed in simulations \citep{Seo2019a}, and is plausible for the Milky Way (Sect.~\ref{subsec:cmz}).

Nuclear disks and bars follow well-defined scaling relations, with mass correlating with size; in particular, nuclear disks typically extend to about 10\% of the bar length \citep{Gadotti2025a}. Many galaxies host both nuclear disks and NSCs, however the scaling relations of nuclear disks are only marginally compatible with those of NSCs. There is no strong observational evidence or theoretical support for a shared relation, as the masses of both components scale with galaxy mass but with different slopes. Furthermore, the peak frequency of NDRs and NSCs occurs in galaxies with different masses, respectively at $\sim 10^{10}$ and $\sim 10^{9}$\Msun \citep{Georgiev2016, Neumayer2020, Gadotti2025a}. 

\subsection{Circumnuclear Rings}
\label{subsec:circumnuclear_rings}

The term ``circumnuclear ring" (or simply ``nuclear ring") refers to ring-like structures commonly found in the centers of barred galaxies \cite[e.g.,][]{Comeron2010}. Circumnuclear rings have diameters ranging from $\sim100-1000$~pc \citep[e.g.,][]{1996FCPh...17...95B,2010MNRAS.402.2462C}. The ring's radius is set by a number of factors including the bar's pattern speed, the circular rotation curve, and the sound speed of the inflowing gas (see \citealt{Sormani2024a} for a brief review). 

Circumnuclear rings are theorized to be formed as a result of the bar orbits and associated resonances \citep[e.g.,][and many others]{Athanassoula1992a,Athanassoula1992b}. As gas flows in towards the galaxy's center, it transitions from the bar's $x_1$ orbits to the $x_2$ orbits. This material will then accumulate at the location of the innermost $x_2$ orbit. In their simulations, \citet{Sormani2024a} found that circumnuclear rings lie at the inner edge of the gap cleared by the Inner Lindblad Resonance (ILR); in other words, circumnuclear rings are inside of the ILR. While this is contrary to previous theories, where the ring is co-located with the ILR \citep[e.g.,][]{1996FCPh...17...95B}, it agrees better with the results of simulations \citep[e.g.,][]{Patsis2000,Li2015,Sormani2015} and observations \citep[e.g.,][]{Levy2022,2023ASPC..534...83H}.

The gaseous material in circumnuclear rings is primarily molecular. As gas accumulates, it is compressed and shocks with new material entering the ring, enabling rapid cooling \citep{Morganti2018} and high molecular gas masses ($\sim10^8-10^9$~\Msun; e.g., \citealt{Querejeta2021}). In addition, circumnuclear rings are usually the sites of the most extreme and active star formation in a galaxy \citep{Moon2021,Levy2022,Schinnerer2023,Sun2024}. Unlike the NSD, the stars formed in a circumnuclear ring form as massive bound star clusters \citep{Krumholz2019}. These young massive star clusters (YMCs) orbit within the ring, with typical orbital timescales of a few Myr \citep[e.g.,][]{Levy2022,Sun2024}. Feedback from these YMCs can power galaxy-scale outflows (without requiring an AGN) where the base of the outflow is set by the size of the circumnuclear ring \citep[e.g.,][]{Krieger2019,Mills2021,Nguyen2022}.

In the Milky Way, the gaseous circumnuclear ring is referred to as the ``Central Molecular Zone" (CMZ) \citep[e.g.,][]{Mills2017,Battersby2020,2023ASPC..534...83H,2025ApJ...984..156B} and has a diameter of $\simeq240$~pc \citep{2023ASPC..534...83H}. The CMZ hosts the largest and densest concentration of molecular gas, the most massive and actively star-forming regions, and two most massive YMCs, Arches and Quintuplet, in the Galaxy \citep[e.g.,][]{Battersby2020,Hosek2022,2023ASPC..534...83H,Longmore2026}. Additional information about the CMZ can be found in Sect.~\ref{subsec:cmz}.

Outside of the Milky Way, the nearby ($d=3.5$~Mpc; \citealt{Rekola2005,Congiu2025}) starburst galaxy NGC\,253 is one of the best-studied examples of galaxies hosting a prominent circumnuclear ring. 
NGC\,253 is a roughly Milky Way mass galaxy \citep{Leroy2021} viewed nearly edge-on ($i\approx76^\circ$; \citealt{McCormick2013}). The central $\sim$kiloparsec has a star formation rate $\approx2$~\Msun~yr$^{-1}$ \citep{Bendo2015,Leroy2015}, and this nuclear starburst is powering a large-scale multiphase outflow along the galaxy's minor axis \citep[see e.g.,][and references within]{Lopez2023,Cronin2025}. The immense dust columns to the center of this galaxy ($A_V\approx17-4200$; \citealt{Kornei2009,Leroy2018}) mean that the nuclear stellar structures are only visible at infrared wavelengths and longer. In practice, this limits us to observations of very recent star formation ($t\lesssim6$~Myr), where the natal molecular gas and dust has not yet been completely dispersed \citep{Sun2024,Schinnerer2024}. In NGC\,253, the vast majority of the recent star formation in this galaxy is concentrated in $>14$ YMCs \citep{Leroy2018,Levy2021}. Moreover, the architecture of these YMCs (and the dense molecular gas) are morpho-kinematically consistent with a circumnuclear ring with a diameter of $\approx220$~pc \citep{Levy2022}. Ages are challenging to determine for these objects, but a lack of synchrotron emission in their radio spectral energy distributions suggests a lack of supernovae and hence ages $\lesssim3$~Myr \citep[][see also \citealt{Rico-Villas2020}]{Mills2021}. Several of these embedded YMCs show signs of strong stellar feedback (\citealt{Levy2021}, in prep.; Sheriff et al., in prep.), which may contribute to the large-scale galactic outflow \citep{Mills2021}.

\subsection{Inner stellar bars}
\label{subsec:nuclear_stellar_bars}

The term ``inner bar'', sometimes known as ``nuclear bar", refers to a second stellar-bar structure embedded in the center of a barred galaxy. A high fraction of disc galaxies host double bars, with values depending on galaxy mass and Hubble type. On average, circa 50$\%$ of  massive ($\log_{10}(M_{\star}/M_{\odot}) > 10.5$) barred galaxies are actually double barred \citep[][]{Erwin2004b,Erwin2024a}. While inner bars are considered ``small systems" due to their coexistence with regular, larger bars, once isolated they can be as long as single bars in other galaxies. Bar strength is related to the effect that these elongated structures have on the gravitational potential of the galaxy. Inner bars tend to be weaker than single or outer bars \citep{deLorenzoCaceres2019b, Erwin2019}.

Since the first identification of NGC~1291 as a double-barred galaxy by \citet{deVaucouleurs1975}, much work has been devoted to understanding the formation of two coexisting non-axisymmetric structures within axisymmetric disks, both numerically and observationally \citep[][among others]{FriedliAndMartinet1993, Heller2001, EnglmaierAndShlosman2004,Wozniak2015}. The TIMER survey (Sect.~\ref{subsec:nuclear_stellar_disks}) allowed determining that some inner bars are dynamically formed out of nuclear stellar disks \citep{deLorenzoCaceres2019a,Bittner2020a}. Such a process is fully analogous to the dynamical formation of regular stellar bars in the main galaxy disks \citep[e.g.][]{DebattistaAndSellwood2000}. Notwithstanding, some bimodality in the photometric properties of inner bars has been found \citep{deLorenzoCaceres2020}, suggestive of a second formation channel for part of the population of double bars. Recent numerical work has confirmed that tidal interactions are a feasible way of forming stable inner bars \citep{Semczuk2024}.

Deriving dynamical ages for stellar structures is a very challenging task, as there are no star formation events matching with the stellar redistribution. The newly developed methodology presented in \citet[][]{deSaFreitas2023a,deSaFreitas2025a} for inferring the dynamical formation moment of regular bars cannot be applied to inner bars, since these do not form nuclear stellar disks inside them. Despite of this fact, the ages of a handful of inner bars have been constrained through star formation history measurements, finding always medium-age systems. Inner bars are younger than their corresponding outer bars \citep{deLorenzoCaceres2013,deLorenzoCaceres2019a}. More measurements are needed to complete our knowledge in this regard.

Observations have shown that inner bars are long-lived structures, are able to buckle and develop boxy/peanut structures as regular bars do \citep[][]{MendezAbreu2019a}. Moreover, simulations have demonstrated that a double-barred system \citep{Shlosman1989}, or even just a small-scale bar \citep[alike an inner bar,][]{Du2017a}, is capable of transporting gas to the very central regions of galaxies, further inwards than allowed by a regular-sized single bar. The role of inner bars in funneling gas to the central parsecs and thus feeding MBHs, or even in the formation of some NSCs, are matter of debate. 

\subsection{Nuclear star clusters}
\label{subsec:nuclear_star_clusters}

NSCs are the most frequently observed stellar structures in galactic nuclei. They have been found to coexist with MBHs, but there are also numerous NSCs without an observationally confirmed central BH 
\citep{Neumayer2020}.
NSCs share some properties with globular clusters, such as similar sizes and, in some cases, overlapping masses. However, NSCs can reach higher masses and central densities, making them promising environments in which BHs may form, be retained, and potentially grow over time.

\begin{figure*}
\centering
\includegraphics[width=0.7\textwidth]{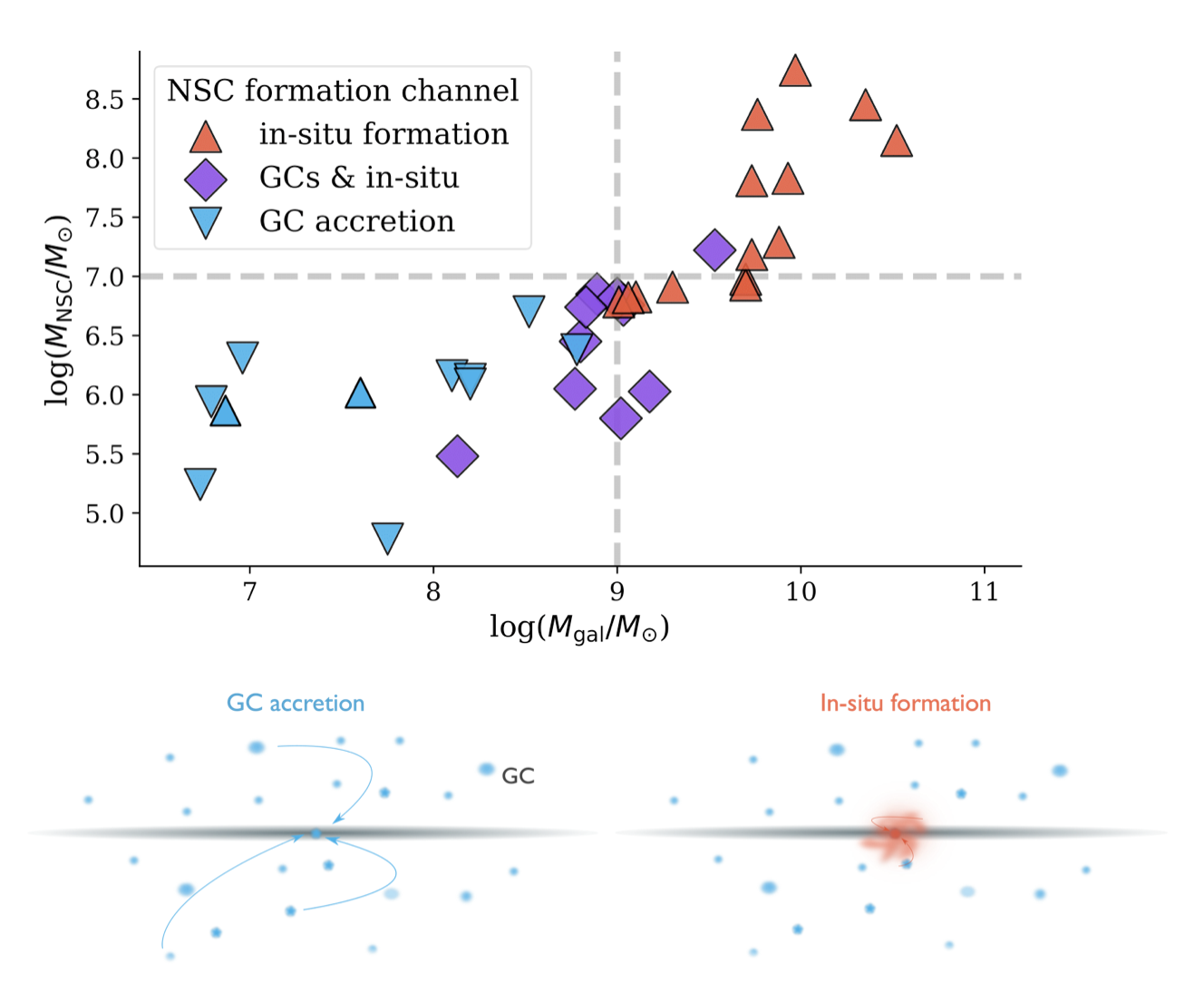}
\caption{Schematic overview of the main formation channels of nuclear star clusters (NSCs). For low-mass galaxies, NSCs have likely formed from globular clusters that spiralled into the galaxy center due to dynamical friction. On the other hand, NSCs in massive galaxies are dominated by the in-situ formation scenario. Here, the nuclear star cluster formed directly at the center out of star-forming gas. Adapted figures from Katja Fahrion's personal webpage (\url{https://katjafahrion.github.io/}).}
\label{fig:nscformation}
\end{figure*}

\subsubsection{Occupation statistics}

NSCs are very common in the universe, being hosted by more than 77\% of late-type galaxies (LTGs) and more than 66\% of early-type galaxies (ETGs, see review by \citealt{Neumayer2020}). 
The occupation fraction of NSCs in dwarf galaxies is sensitive to the environment of their host, first discussed in \citet{SanchezJanssen2019a}: 
galaxies in dense galaxy clusters have a higher probability of hosting a NSC at a specific galaxy mass than their counterparts in looser galaxy clusters, galaxy groups, or voids.
As discussed most recently in \citet{Zanatta2024a}, properties of functional fits to the observed occupation fraction relate to the halo mass of the galaxy's local environment, hinting at a deeper connection between the formation of the host galaxy and its NSCs.
At the massive end, the nucleation fraction drops relative to the peak at intermediate masses, and galaxies increasingly tend to host MBHs, with BH masses increasing as a function of galaxy mass in the case of massive elliptical galaxies. 
The most massive ellipticals have cored light profiles \cite[e.g.,][]{Thomas2014} and apparently do not host any NSCs (\citealt{Neumayer2020, Graham2012}, but see also Sect.~\ref{sub:nsc_formation}). 



\subsubsection{Formation Channels} \label{sub:nsc_formation}
As mentioned in Sect.~\ref{sec:intro} and shown in Figure \ref{fig:nscformation}, the formation pathways of NSCs split into two categories: gas accretion followed by in-situ star formation \citep[e.g., ][]{Loose1982, Milosavljevic2004, Pflamm2009}, or accretion and merger of star clusters \citep[e.g., ][]{Tremaine1975, Capuzzo-Dolcetta1993, Portegies2006, Capuzzo-Dolcetta2008a, Capuzzo-Dolcetta2008b, Antonini2012}, while hybrid scenarios have also been proposed \citep[e.g., ][]{Hartmann2011, Antonini2013, Guillard2016}.
In the first case, globular clusters experience dynamical friction-induced migration towards the galaxy center where they may merge and combine to form the nuclear star cluster if core-stalling is unimportant.
In an alternate pathway, a nuclear star cluster may be accreted during a galaxy merger event and can, depending on the conditions of the galaxy merger, reach the galaxy center within a Hubble time.
The second formation scenario requires gas migration and subsequent star formation in galaxy centers.
Star formation alone is not sufficient, however, because newly formed stars must form a bound star cluster that resists strong tidal forces in a galaxy center region \citep[][]{Portegies-Zwart2002a}.

These two formation pathways (star-cluster migration and \textit{in-situ} star formation) are not mutually exclusive, and both are likely to contribute to the build-up of nuclear star clusters over a broad range of galaxy masses \citep[e.g.][]{Fahrion2021a}.
Their relative contribution depends on the galaxy type. Globular cluster infall appears to be the preferred path for low-mass galaxies, while in situ formation may be the dominant path in more massive ETGs and in spiral galaxies \citep{Pinna2021, Fahrion2022a}.
It remains unclear if an NSC can exclusively form \textit{in-situ} or from migrating star clusters.

Massive galaxies often lack NSCs, in particular elliptical galaxies. This is often attributed to their disruption by the erosion from two (or more) MBHs that were brought to the galaxy center through dynamical friction following a major galaxy merger \citep[e.g.][]{Antonini2019a}.
Support for this idea comes from the result that the occupation fraction appears to remain constant (or, at least, does not decrease) for spiral galaxies above $10^{10} \, \mathrm{M}_{\odot}$ that are believed not to have experienced recent major mergers \citep{Ashok2023a}.
At the same time, migration of globular clusters towards a galaxy center may become inefficient in more massive galaxies, both due to the increasing dynamical friction time-scales and mass-loss from the star-cluster denser environment, causing a depletion of massive star clusters in galaxy centers \citep[][]{Leaman2022a}. 
To date, the dominant mechanism remains unclear.

\subsubsection{Modeling of NSC formation}  \label{sub:nsc_simul}
NSC formation can be modeled with different approaches corresponding to different types of simulations:
\begin{itemize}
    \item \textit{Direct $N$-body simulations}: These are 
    simulations in which the 
    evolution of all particles is followed directly, making them well suited to study two-body relaxation, close encounters, tidal effects, and other dynamical processes in dense stellar systems \citep[e.g.,][]{Antonini2012}. They have been widely used to investigate NSC formation through the inspiral and merging of star clusters \citep{Capuzzo-Dolcetta1993, Baumgardt2003, Antonini2012, Mastrobuono-Battisti2014, Arca-Sedda2015, Tsatsi2017, Askar2021}.  Merger models tailored to the Milky Way can reproduce several key properties of the Galactic NSC, including its mass, density profile, flattening, and rotation, supporting globular-cluster inspiral as an important contributor to its assembly (\citealt{Antonini2012, Tsatsi2017}. However, observational studies based on NSC metallicity rule out a major contribution from this scenario, see Sect.\ref{subsec:cmz}). The high stellar densities reached in these systems also make stellar collisions and close encounters dynamically relevant, especially in the innermost regions \citep{Freitag2006b, Mastrobuono-Battisti2021b}. Their main limitation is computational cost: despite major progress enabled by GRAPE hardware and, more recently, GPUs, direct $N$-body simulations still generally fall short of modelling NSCs with their full physical number of stars, especially in the most massive systems.
    \item \textit{Hydrodynamical models}: These models include stars and gas and combine N-body simulations (stellar dynamics) and hydrodynamics (gas physics) to allow for a multi-phase NSC formation \citep[e.g., ][]{Lucas2013}. These simulations allow us to trace both dissipative and dissipationless processes. They also allow combinations and synergy between cluster mergers and in-situ formation \citep[e.g.][]{Guillard2016}. However, they are computationally expensive, making it complex to calibrate feedback. They have revealed that NSCs can form through joint gas-driven and dynamical processes, varying with galaxy mass and merger history \citep{Guillard2016}. 
    \item \textit{Semi-analytic models}: These models combine analytical prescriptions for processes such as dynamical friction, cluster disruption, gas inflows, and star formation. They are computationally efficient and therefore well suited to statistical studies of large galaxy populations \citep[e.g.,][]{Gnedin2014}. In some cases, they are coupled to galaxy evolution models in a cosmological framework, allowing one to investigate the relative roles of cluster migration and \emph{in situ} star formation in shaping NSC--host scaling relations \citep{Antonini2013,Antonini2015}. Their main limitation is that they do not directly resolve collisional dynamics, close encounters, or relaxation processes, and therefore rely on approximate treatments of the underlying physics \citep{Agarwal2011,ArcaSedda2014,Askar2022}.
   \item \textit{Cosmological simulations}:
   This option provides a cosmological context at the expense of spatial and mass resolution. Two approaches are common to address this issue:
    \begin{itemize}
        \item NSCs embedded in a galaxy assembly
    context (the NSC is placed there)
        \item Zoom-in approach: this allows to resolve NSCs + MBHs \citep[e.g.,][]{Gray2025}
    \end{itemize}
    Among the most important insights, it has been shown that gas-rich mergers lead to nuclear inflows and in-situ star formation \citep{Ma2020, Hopkins2010b}, and how NSC can form from hierarchical galaxy growth \citep{Pfeffer2014, Biernacki2017} or through cluster mergers \citep{vanDonkelaar2024}. 
    With the strength of putting NSCs in the cosmological context and connecting them to the cosmic environment, its clear limitations are resolution challenges, and the sub-grid physics for star formation and feedback, which are still far from being fully realistic. 
\end{itemize}

These are the main take-home messages from simulations of NSC formation:
\begin{itemize}
    \item The structural properties of the Galactic NSC can be explained if it formed through the infall and merger of massive and dense stellar clusters.
    \item In situ star formation is necessary.
    \item Different methods should be combined to understand how NSCs form.
    \item NSC structure and kinematics can indirectly reveal MBH binaries.
\end{itemize}

\section{Central black holes}
\label{subsec:central_black_holes}

MBHs in galactic nuclei, with masses typically in the range
$10^{4}$--$10^{10}\,\mathrm{M_{\odot}}$, are key components of galaxy evolution.
During active accretion phases, they can inject energy into the interstellar and
circumgalactic medium, regulating star formation and shaping the structure of
the host galaxy's disk and halo. As a result, galaxies and their central MBHs
co-evolve, producing scaling relations between MBH mass and host-galaxy
properties such as the bulge mass \citep[e.g.][]{Haering2004a} or the stellar
velocity dispersion \citep[e.g.,][]{Gebhardt2000,FerrareseMerrit2000}. 
MBHs and NSCs are also expected to co-evolve with their host galaxies, reflecting the coupled growth of BHs, stellar nuclei, and galaxy-scale properties \citep{Antonini2015}.

\subsection{Occupation fraction}

What fraction of galactic nuclei harbor an MBH remains one of the central open
questions, particularly at the low-mass end of the galaxy population. For the
most massive elliptical galaxies it is well established that they host central
MBHs \citep[e.g.][]{Kormendy2013}, but the situation is less clear for massive
spirals and for lower-mass galaxies \citep[][]{Graham2009,Greene2020a,Zou2025}.
Constraints from nuclear X-ray emission, direct dynamical measurements, and
recent multi-wavelength studies all suggest a decline in occupation fraction
toward lower galaxy masses \citep[][]{Miller2015a, Thater2017,Nguyen2019b,Burke2025,Zou2025}.
In particular, \citet{Burke2025} derived lower limits of $\gtrsim 90\%$ at
$M_{\star}=10^{8}\,\mathrm{M_{\odot}}$ and $\gtrsim 39\%$ at
$M_{\star}=10^{7}\,\mathrm{M_{\odot}}$, while \citet{Zou2025} found a decline
from $>93\%$ in massive galaxies to $66^{+8}_{-7}\%$ at
$M_{\star}\sim10^{9}$--$10^{10}\,\mathrm{M_{\odot}}$ and to $33^{+13}_{-9}\%$ at
$M_{\star}\sim10^{8}$--$10^{9}\,\mathrm{M_{\odot}}$. On the theoretical side,
models in which seed formation and subsequent MBH growth depend on halo mass,
gas supply, and environment likewise predict a declining occupation fraction
toward lower galaxy stellar masses, although the precise form of this decline depends
sensitively on the adopted seeding prescription
\citep[e.g.][]{Ricarte2018b,Greene2020a,Tremmel2024,Bhowmick2025}. The observed range likely reflects both genuine variation in MBH seeding and subsequent growth, and the challenge of detecting low-mass, low-luminosity MBHs \citep[e.g.][]{Greene2020a,Graham2021b}.
Tidal disruption events (TDEs) provide
a complementary probe of otherwise quiescent MBHs, and ongoing time-domain surveys are expected to expand this sample significantly
\citep{Bricman2020}. The occupation fraction at low galaxy masses therefore remains a key local constraint on MBH seed formation and early growth.

\begin{figure*}
\centering
\includegraphics[width=1\textwidth]{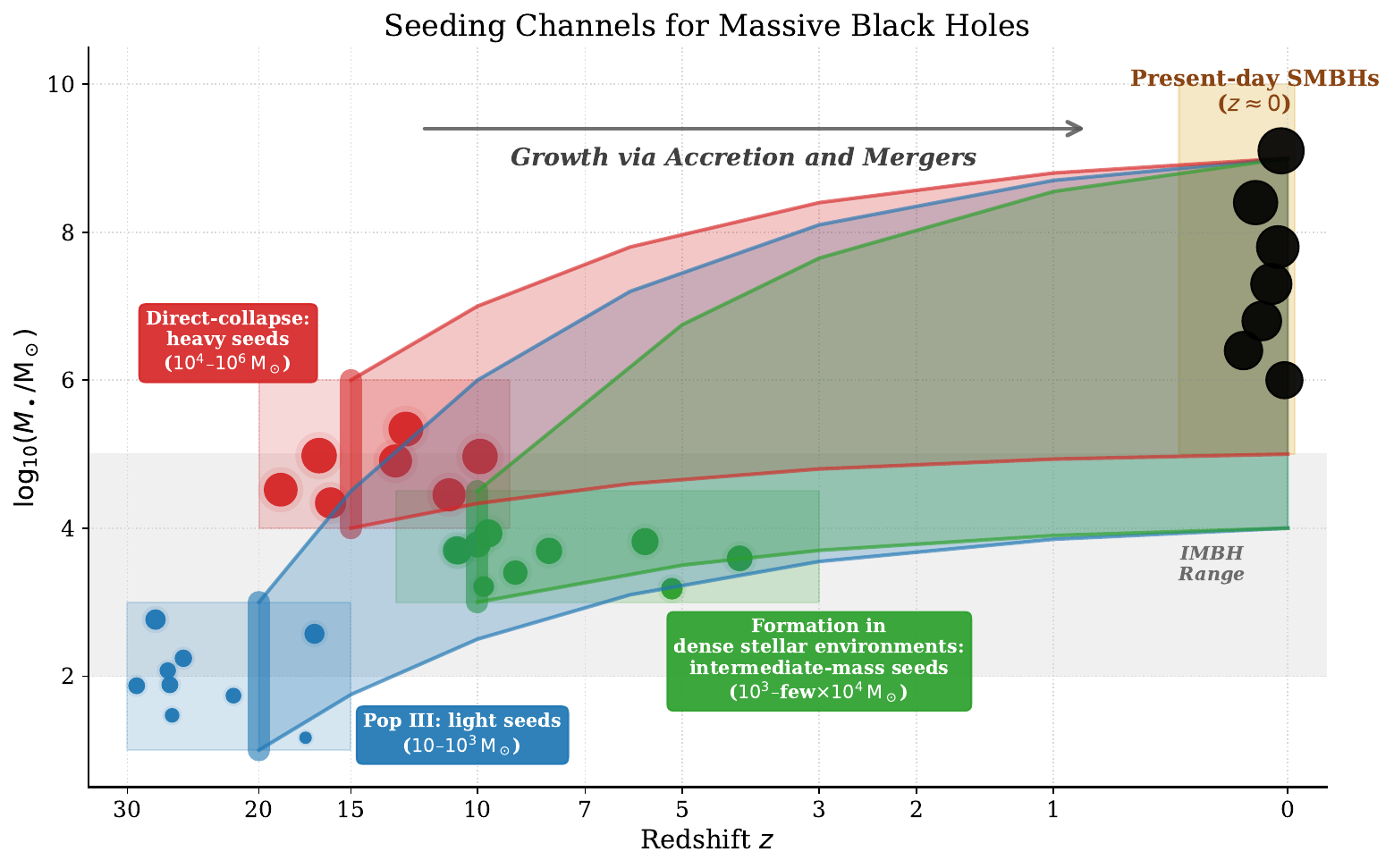}
\caption{Schematic overview of the main seeding channels proposed for MBHs as a function of BH mass and redshift, including light seeds from Population~III stellar remnants, intermediate-mass seeds formed in dense stellar environments through multiple pathways (see \citet{Askar2023} and references therein), and heavy seeds from direct collapse. The shaded tracks illustrate possible growth through accretion and mergers toward the MBHs observed in the present-day Universe, while the grey band marks the IMBH mass range (taken here to be $10^2$--$10^{5}\,\mathrm{M_{\odot}}$). The individual seed and present-day MBH masses shown are illustrative and randomly sampled within their respective mass ranges. Adapted from Fig.~1 of \citet{Mezcua2017-imbh}, Fig.~1 of \citet{Greene2020a} and Fig.~22 of \citet{LISA2023}.}
\label{fig:mbh_seeding_channels}
\end{figure*}

Some relatively nearby galaxies appear to lack central MBHs, or host one below
current detection limits; well-known examples include M~33
\citep[][]{Gebhardt2001,Merritt2001} and NGC{\,}300 \citep[][]{Neumayer2012}.
Their apparent absence may have several origins: seeding may have been
inefficient, a seed may have formed but failed to grow, or an early MBH may
have been displaced or ejected either by gravitational-wave recoil following
MBH mergers or by strong few-body interactions \citep{Hoffman2007,Antonini2015,Askar2021,vanDonkelaar2025}. Such
processes can produce wandering MBHs outside galactic nuclei, although
individual candidates remain debated \citep[e.g.][]{Van-dokkum2025c,Sanchez-Almeida2026}.
In other cases, the MBH and the surrounding stellar nucleus may have evolved
partly independently. Henize~2-10 is often highlighted as an example in which
the central MBH appears offset from the youngest stellar component, suggesting
that the formation of the MBH precedes that of a dominant NSC
\citep{Reines2011,Nguyen2014}. Related observations also suggest that MBH
seeding and growth in galactic nuclei may in some cases be linked to dense
stellar systems and infalling clusters \citep{Graham2021,Graham2021b}.

\subsection{Seeding and Growing MBHs}

The seeds of central MBHs can originate through several channels (see Fig.\,\ref{fig:mbh_seeding_channels}), whose
relative importance likely varies with galaxy mass and cosmic epoch \citep[][and references therein]{Volonteri2012,Greene2020a,Izquierdo-Villalba2023}. In the conventional light-seed channel,
remnants of massive Population~III stars ($\sim 10$--$10^{3}\,\mathrm{M_{\odot}}$)
must grow, in at least some cases at super-Eddington rates, to reach MBH
masses by $z\sim 0$ \citep{Mehta2024}. Heavy seeds instead form more directly through the collapse
of primordial gas clouds or quasi-star--like objects in metal-poor
atomic-cooling halos, producing initial masses of $10^{4}$--$10^{5}\,\mathrm{M_{\odot}}$.
The existence of luminous quasars already at $z\gtrsim 6$ \citep[e.g.][]{banados2018a,2024MNRAS.527.1033B,2024Natur.636..594J}
indicates that heavy seeds, rapid early growth, or a combination of the two must
operate in at least a fraction of systems \citep{Spinoso2023,Regan2024}. A more speculative possibility is that
some MBH seeds are primordial BHs that formed in the very early Universe \citep[][and references therein]{Prole2025,Ziparo2025}.

Dense stellar environments provide another important pathway for MBH seeding.
As reviewed by \citet{Askar2023} and references therein, IMBH seeds may form
through runaway stellar collisions \citep{spz2002a,SPZ2004}, repeated or hierarchical
mergers of stellar-mass BHs \citep{Miller2002,Antonini2019a,Gerosa2021,Atallah2023}, and related dynamical \citep{Rizzuto2021,DiCarlo2021,gonz2021} and gas-mediated processes \citep{Vesperini2010,Davies2011,Leigh2013,Natarajan2021,Das2021} in environments such as dense globular clusters, NSCs, and Population~III star clusters \citep{Reinoso2025}. 

These intermediate-mass seeds may be transported to galactic nuclei via the inspiral of star clusters driven by dynamical friction, contributing to the assembly of NSCs and the growth of central MBHs \citep{Ebisuzaki2001,SPZ2006b,ArcaSedda2018,Askar2022,Fragione2022}. Once in the nuclei, these seeds may grow substantially through gas accretion, mergers with other BHs, and the tidal capture and disruption of stars \citep{stone2017,Ryu2024}.
Recent works have strengthened the case for this broad
stellar-dynamical channel, showing efficient collisions and IMBH formation in compact
clusters over a wide range of initial conditions and metallicities
\citep{Giersz2015,Mapelli2016,Sakurai2017,Fujii2024,Rantala2025,Vergara2025a,Vergara2025b}. In this
sense, dense stellar systems may provide a natural bridge between stellar-mass
BHs and the low-mass end of the nuclear MBH population, while enhanced
X-ray detection rates in high-dispersion NSCs further support the importance of
such environments for MBH formation and growth \citep{Baldassare2022}.

A growing number of observed systems reinforces the connection between dense
stellar systems, stripped galactic nuclei, and IMBHs. \citet{Haeberle2024b} reported seven fast-moving stars within the central $3\arcsec$ of the largest MW globular cluster 
$\omega$~Centauri, implying a firm lower limit of
$\sim 8200\,\mathrm{M_{\odot}}$ on the mass of an IMBH.
Since $\omega$~Centauri is widely regarded as the stripped nucleus of an accreted
dwarf galaxy, it provides a local link between massive star clusters, stripped
nuclei, and MBHs. 
Dynamical modelling of high-resolution integral-field spectroscopy of
B023-G078, the most massive globular cluster in M31 and another
stripped-nucleus candidate, infers the presence of a central BH
with $M_{\bullet}\sim 10^{5}\,\mathrm{M_{\odot}}$ \citep[][]{Pechetti2022a}.
These cases underline a broader observational point:
IMBHs, and the low-mass end of the MBH population, can remain undetected
unless observations have sufficient spatial resolution and sensitivity to
probe stellar and gaseous motion within the BH's sphere of influence \cite{Mezcua2017-imbh}.

NSCs likely play a dual role: they can provide favourable conditions for seed
formation via stellar collisions, and they can enhance subsequent MBH growth by
funnelling gas into the central few parsecs. Semi-analytic models that include
seed formation through stellar collisions in NSCs reproduce a broad MBH mass
function, and predict that more compact NSCs are the most efficient sites of
seed production \citep{Liempi2025,Paiella2026}. High-resolution simulations of low-mass
galaxies further indicate that the presence of an NSC can enhance both nuclear
star formation and MBH accretion \citep{Shin2025}, whereas MBHs in systems lacking NSCs tend to
grow inefficiently \citep{Partmann2024}. The occupation fraction of MBHs in
low-mass galaxies is therefore shaped by a combination of seed formation, early
cluster and nuclear dynamics, later gas supply, and subsequent mergers and
recoil events.

\subsection{Active Galactic Nuclei (AGN)}

The traditional view of galaxy evolution has long attributed the regulation of
star formation in low-mass galaxies primarily to stellar feedback
\citep[e.g.][]{Dekel&Silk1986,Hopkins2014}. However, AGN feedback is
increasingly recognized as an additional ingredient in MBH--galaxy
co-evolution in this regime. Theoretical studies by \citet{Silk2017} and
\citet{Dashyan2018} argued that AGN may, in some cases, be more effective than
stellar feedback at clearing gas from low-mass galaxies, and the growing census of AGN in dwarf galaxies provides direct evidence that at least a subset of these systems hosts accreting MBHs
\citep{Reines2013,Mezcua2018,Mezcua2020HiddenAGN,Pacucci2021,Reines2022,Graham2021b}.
Using DESI early data, \citet{Pucha2025} identified 2444 AGN candidates in dwarf
galaxies, more than tripling the known census, and reported an observed
dwarf-galaxy AGN fraction of $\sim 2.1\%$. The observed AGN fraction, however,
should not be conflated with the true MBH occupation fraction, since only a
subset of MBHs are actively accreting at any given time.

Cosmological simulations still give a diverse picture for the growth of MBHs
and the impact of AGN feedback in low-mass galaxies. Some models, such as
\texttt{EAGLE} and \texttt{FABLE}, suggest that stellar feedback hinders
MBH growth by starving the central accretion flow
\citep{Dubois2015,Koudmani2021ASimulations}, whereas others, such as
\texttt{ROMULUS} and \texttt{IllustrisTNG}, produce a substantial population of
MBHs in low-mass systems
\citep{Sharma2020BlackGalaxies,Haidar2022TheSimulations}. High-resolution
zoom-in simulations from the \texttt{AURIGA} project show that in the case when MBHs in
dwarf galaxies grow, they establish scaling relations consistent with local
observations and suppress star formation primarily through episodic heating
rather than complete gas removal \citep{ArjonaGalvez2024}. In some systems this
may lead to a dual-mode cycle in which AGN-driven heating and outflows suppress
star formation on galactic scales, while shocks and compression locally enhance
or sustain star formation in the nuclear region
\citep[e.g.][]{MercedesFeliz2023a,CresciMaiolino2018,Combes2017}. The new
\texttt{COLIBRE} simulations, a successor to \texttt{EAGLE}, underline that
the outcome depends sensitively on the AGN feedback prescription: in addition
to a fiducial thermal model, they include a spin-dependent hybrid jet/thermal
scheme and find that AGN feedback can affect galaxies down to
$M_{\star}\sim 10^{8}\,\mathrm{M_{\odot}}$
\citep{Schaye2025a,Husko2025a,Chaikin2025a}. Predictions from different
simulations diverge most strongly at high redshift \citep{Habouzit2022b},
highlighting the need for further calibration against observations.

Several lines of evidence indicate that AGN activity is an episodic and
recurring phase in the lives of many galaxies, rather than a permanent
property of a special class. The ubiquity of MBHs in local bulges implies
that most massive galaxies must have undergone at least one accretion episode
in the past \citep{Page2004}. The AGN fraction depends on host-galaxy
properties and redshift in a way consistent with a transient phase
\citep{Shabala2008,Wang2017}, and inferred duty cycles can be as short as
$10^{4}$--$10^{5}\,$yr in some systems \citep{Morganti2017}. The connection
between AGN activity and circumnuclear star formation also suggests that
nuclear fuelling, starbursts, and feedback may often form part of an
evolutionary sequence rather than representing fully independent phases
\citep[e.g.][]{Page2004,Storchi-Bergmann2019}. Lastly, over the past decade, a growing number of so-called changing-look AGN have been observed to show very strong flux and spectral changes \citep[e.g. review by][]{Ricci2023}. While some changing-look AGN can be explained by clouds or outflows eclipsing the central engine of the AGN, others are likely triggered by strong changes in the accretion rate of the MBH. In this view, every galactic
nucleus hosting an MBH is a potential AGN, and whether it appears active
depends on its current gas supply, accretion rate, and the timescale on which
it is observed.

No single mechanism triggers all AGN activity: the dominant process depends on AGN
luminosity, host-galaxy properties, gas supply, and redshift \citep[see the
review by][]{Storchi-Bergmann2019}. Galaxy mergers are thought to be important
for the most luminous AGN \citep{LaMarca2024,Satyapal2014,Li2023}, whereas
secular and stellar-driven processes may dominate for lower-luminosity systems
\citep{Cisternas2011,Garland2024,daSilva2024}. In dwarf galaxies, recent MaNGA
results indicate that mergers can enhance the AGN fraction at projected
separations below $\sim 20\,$kpc, especially after first passage, but are not
the dominant triggering mechanism overall \citep{Erostegui2025}. Gas inflows
to the MBH can also be driven by nuclear instabilities, often facilitated by
bars, as seen in both observations and simulations
\citep[e.g.,][]{Audibert2019a,Combes2023a,Winkel2025b,MercedesFeliz2023a}.

AGN feedback can be both negative and positive, and can operate on both
nuclear and galactic scales. On nuclear scales, observations suggest
alternating ``build-up'' and ``destruction'' phases, in which gas accumulates
and is then expelled once the AGN reaches sufficient power
\citep[][]{Shin2019a,Bessiere2022a,GarciaBurillo2024a}. On larger scales,
outflows, turbulence, and heating can suppress star formation, while shocks
and compression may locally enhance it
\citep{CresciMaiolino2018,Hopkins2016,Silk2013,Zubovas2013,Combes2017}. In
low-mass galaxies, the repeated redistribution of gas may even help lower the
central dark-matter density, contributing to the transformation of cusps into
cores \citep{ArjonaGalvez2024,Pontzen2012,DiCintio2014}.

Identifying AGN in dwarf galaxies remains observationally challenging: their
low metallicities often move them into the star-forming or composite regions
of standard emission-line diagnostic diagrams \citep{Trump2015,Cann2019}.
Even so, recent IFU studies show that dwarf-galaxy AGN can drive measurable
outflows. In the MaNGA AGN dwarf galaxies survey, \citet{RodriguezMorales2025}
identified 11 dwarf galaxies with broad [O\,{\sc iii}] components and outflow
signatures, with velocities of $W_{80}\sim 205$--$566\,\mathrm{km\,s^{-1}}$,
indicating that AGN feedback in this regime can be kinematically important.

A recent step forward has come from \jwst{}, which is opening a new window on
low-luminosity AGN (LLAGN). The ReveaLLAGN programme (``Revealing
Low-Luminosity Active Galactic Nuclei'') uses \jwst{}/NIRSpec and MIRI to
isolate nuclear emission in nearby LLAGN that had previously been difficult
to separate from host-galaxy light \citep{Seth2021a,Goold2024a}. The first
survey results show that \jwst{} can detect high-ionization lines far fainter
than previously accessible, and reveal a transition near
$\log(L_{\mathrm{bol}}/L_{\mathrm{Edd}})\sim -3.5$, below which LLAGN
spectral energy distributions become increasingly deficient in ultraviolet
photons \citep{Goold2026}. Combined with improved constraints on the MBH
occupation fraction, such observations are essential for connecting the
demographics of dormant and active MBHs in galactic nuclei and for refining
predictions for future electromagnetic and gravitational-wave surveys.

\section{The central region of our Galaxy} 
\label{subsec:cmz}
Our Galaxy, the Milky Way, contains several of the previously discussed structures: an MBH, known as Sgr~A*, an NSC, an NSD, and a ring-shaped complex of dense molecular gas, called the central molecular zone (CMZ). 

The mass of Sgr~A* is known with unequaled high precision ($M_\bullet$=(4.30$\pm$ 0.012)$\cdot$10$^6$\,$M_\odot$, \citealt{2022A&A...657L..12G}). This precision is reached thanks to long-time monitoring of stars orbiting Sgr~A* \citep[e.g.][]{2016ApJ...830...17B,2017ApJ...837...30G,2019Sci...365..664D}, and fitting the stellar orbits of the so-called S-stars. Sgr~A* has shown overall low luminosities in the decades after its discovery \citep[e.g.][]{2025arXiv250320081C}, though there is some variability in brightness \citep[e.g.][]{2019ApJ...882L..27D}. However, X-ray reflection observations of CMZ clouds indicate that there has been energetic activity of Sgr~A* a few hundred years ago \citep[e.g.][]{2010ApJ...714..732P,2022MNRAS.509.6068K}. 

The Milky Way NSC has an effective radius of $\sim$4-7\,pc and a mild flattening (minor-to-major axis ratio) $q\sim$0.7 \citep{2014A&A...566A..47S,2016ApJ...821...44F,2020A&A...634A..71G,2026ApJ..1002...71V}. 
The mass of the NSC is $\sim$2--4 $\times 10^7$~\Msun~ \citep{2014A&A...566A..47S,2015MNRAS.447..948C,feldmeierkrause2017,2025A&A...699A.239F,feldmeierkrause2026,2026ApJ..1002...71V}. This extremely dense star cluster rotates with a velocity of $\sim$20--40\,\kms \citep{2025A&A...699A.239F,feldmeierkrause2026,2026ApJ..1002...71V}. 
The formation mechanisms responsible for the mass assembly of the Milky Way NSC are not entirely clear, and its complex nature suggests a complex formation and evolution. A kinematic complexity includes a misalignment with the galaxy's main disk and an additional kinematically decoupled inner component, suggesting different episodes of star–cluster accretion \citep{2014A&A...570A...2F, feldmeierkrause2017}. However, the spread in metallicities suggests a formation in different episodes. While metal-poor stars might come from migrating globular clusters \citep{Feldmeier2017a, Feldmeier2020}, the NSC has relatively high metallicities and is poor in RR Lyrae stars, ruling out globular cluster accretion as the predominant formation mechanism \citep{Dong2017, Schoedel2020, Nandakumar2025}. 

The NSD is a flattened ($q\sim$0.3), dense and rotating stellar structure that dominates the gravitational field of the Milky Way at a Galactocentric radius of approximately 30\textless$R$\textless300 pc. It has a scale radius of  $\sim$100\,pc, a scale height of $\sim$30\,pc \citep{1999A&A...348..768P,2002A&A...384..112L,2022MNRAS.512.1857S}, and a mass of $\sim$1$\times10^9$\,\Msun \citep{2022MNRAS.512.1857S}.
The NSD formed thanks to the Galactic bar transporting large quantities of gas from the Galactic disc at R$\sim$3\,kpc down to the CMZ \citep[e.g.][]{Athanassoula1992b,2014MNRAS.445.3352C,2024MNRAS.528.5742S}, where it produced intense star formation activity. 

The CMZ may be the local analogue of the star-forming gas rings that are commonly found at the centre of external barred galaxies \citep{1996FCPh...17...95B,2010MNRAS.402.2462C}.  This complex of giant molecular clouds has a radius of about 200\,pc \citep{1996ARA&A..34..645M}. The radius of the CMZ gas ring might be slowly growing over Gyr timescales \citep{Seo2019a}, and its star formation activity slowly builds up the NSD inside out (see also Sect.~\ref{subsec:nuclear_stellar_disks}). The total gas mass is estimated to $10^7-10^8$\,\Msun \citep{1996ARA&A..34..645M,Dahmen1998, 2025ApJ...984..156B}.
The gas has high densities, large velocity dispersions, high temperatures and pressures, turbulent motions and strong magnetic fields, which make it the most extreme star-forming environment of our Galaxy and differentiate the CMZ from the star-forming regions in the Galactic disk \citep{Tress2025}. 

Measurements of the Galactic centre stellar structures are hampered by the strong and spatially varying interstellar extinction, restricting existing observations to longer wavelengths $\gtrsim$1.5\,$\mu m$. The visual extinction of $A_V\gtrsim30$ \citep[e.g.][]{2011ApJ...737...73F} makes the Milky Way centre inaccessible at optical wavelengths, and the nuclear structures of the Galaxy are missing in most large Milky Way surveys, including Gaia, Gaia-ESO, RAVE, GALAH, and are only sparsely sampled in APOGEE. 
However, current high-resolution spectroscopy at 8m-class telescopes (e.g., IGRINS@GEMINI) is allowing to obtain detailed abundances of individual stars in the Galactic NSC and NSD, providing insights into the chemical evolution of these structures and their comparison with the inner galaxy \citep{Ryde2025a, Ryde2025b, Nandakumar2025}. This will provide signatures of gas inflow, star-formation efficiency, time delayed enrichment, and constraints on the processes contributing to the formation of these structures. 

Studies have investigated the possibility of a nuclear bar (a ``bar within a bar''-structure) using large photometric catalogues \citep{2001A&A...379L..44A,2011A&A...534L..14G}, to explain how gas can move inwards from the further CMZ. However, no clear detection has been reported yet, and star formation events \citep[e.g.][]{Tress2020b} or magnetic fields \citep[e.g.][]{Tress2024a} might be responsible for inflows. The interstellar extinction and reddening in the Milky Way centre are spatially variable, with several giant molecular clouds along the line-of-sight. This causes large uncertainties in stellar density estimates. It may be possible to detect a nuclear bar from its kinematic signatures in future large-scale kinematic observations \citep{2026arXiv260318738F}. 
Finally, it was reported that star clusters in the CMZ get disrupted quickly, and this may have an impact on the NSC and central stellar structures in the Milky Way \citep[see e.g.][]{PortegiesZwart2001a,PortegiesZwart2002a}.



\section{Stripped galaxy nuclei or clusters}
\label{subsec:stripped_galaxy_nuclei}

During our workshop, we discussed several cases of potentially stripped nuclei or star clusters. We mention here a couple of examples that were presented. 

The oMEGACat catalogue combines high-quality datasets of $\omega${\,}Centauri ($\omega$~Cen) from HST \citep{Haeberle2024a} and MUSE \citep{Nitschai2023a}. 
Current analyses suggest the presence of an IMBH in the center of $\omega$~Cen, with a mass of at least $8200 \, \mathrm{M}_{\odot}$ assuming that the fastest star is bound to the IMBH \citep{Haeberle2024b}. However, a major uncertainty remains the precise determination of the cluster’s center, which critically affects dynamical measurements.
Beyond its internal properties, $\omega$~Cen offers key insights into galaxy assembly. Scaling relations between nuclear star cluster (NSC) mass and host-galaxy mass can be used to estimate the mass of its progenitor galaxy, helping to constrain the accretion event that brought $\omega$~Cen into the Milky Way. In this context, the “$\omega$ dwarf” model (de Souza et al., in prep.) proposes a connection with major accretion structures: Sequoia (outer regions), Gaia-Enceladus (intermediate regions), and $\omega$~Cen itself as the former nucleus.
This scenario is supported by chemo-dynamical analyses based on APOGEE, GALAH, and Gaia data, which reveal consistency between observed trends and model predictions, particularly in the relation between [Ba/Eu] and the galactocentric distance of the progenitor system. 
Finally, recent studies \citep{Clontz2025a, Clontz2025b} provide new constraints on the cluster’s complex stellar populations. They find enhanced helium abundances in second-generation stars, with the degree of enrichment increasing over time and with metallicity. Gaussian mixture modeling of subgiant-branch and red-giant-branch stars indicates that the second stellar population formed roughly 1 Gyr after the first, and exhibits a larger age spread, pointing to an extended and possibly multi-phase star formation history.

ED-2 is a stellar stream that was identified as dynamically different from other stellar structures in Gaia observations of halo stars \citep{Balbinot2023}. 
It is associated with the remnant of a former star cluster (\citealt{Balbinot2024a}, but see also \citealt{Panuzzo2024a}). 
Gaia BH3 is a BH that was identified as part of the ED-2 stellar stream. Its mass places it within the secondary peak of the BH mass distribution inferred from gravitational-wave events, making it particularly relevant for connecting stellar and merger-driven BH populations. 
The progenitor globular cluster of the ED-2 stream would have had a mass in the range $2 \times 10^{3} \lesssim M_{\mathrm{GC}} / \mathrm{M}_{\odot} \lesssim 5.2 \times 10^{4}$ \citep{Balbinot2024a}. However, this estimate depends sensitively on the assumed initial mass function (IMF), with different choices potentially shifting the lower mass limit of the original cluster. 
Today, ED-2 appears to be a highly disrupted system, with only $\sim 25$ stars currently identified. This raises important questions about the completeness of the observed sample and, consequently, about the robustness of the inferred properties of both the remnant cluster and Gaia BH3.

\section{Open Questions and Future Prospects}

We summarize here the key open questions that emerged from the active discussions during our workshop. 

\textbf{Stellar structures}. Regarding the formation of stellar nuclear or inner structures, systematic studies in large samples of galaxies remain lacking. 
This limits the statistical power needed to derive robust scaling relations from unbiased samples. 
For the largest structures, the new Bulge Assembly in Nearby Galaxies (BANG) survey will focus on the central region of a volume-limited sample of 134 galaxies observed with MUSE. This will include the analysis of NDRs in the sample and will provide a more complete census of the central components of galaxies (de Sá-Freitas et al., in prep.).
For smaller structures such as the tiny NSCs, this is more challenging due to the smaller sizes and the overlap with the galaxy complexity at the galactic center, often with additional stellar structures. Furthermore, most of our knowledge about NSCs comes from ETGs, while they are not well studied yet in LTGs. This makes our understanding of NSCs biased and additionally lacking conclusive investigations of trends with morphological types. In the next few years, we aim to cover this gap with new observations, preferably in the infrared spectral range, which allows us to probe dusty stellar nuclei more effectively. 

Another pressing question concerns the role of galaxy interactions in the formation of nuclear structures. During the discussion sessions in our workshop, it emerged that no consensus currently exists. On the one hand, mergers can trigger nuclear starbursts in preexisting central gas reservoirs, and prompt the formation of massive star clusters that can later contribute to the NSC mass assembly, while the accretion of external nuclei and star clusters from the accreted satellite galaxy can favour the inspiral of clusters towards the galactic center (although under debate due to the usual eccentric orbits of accreted clusters). While most galaxies hosting a NSC show signs of past mergers \citep{Pinna2021}, complementing detailed NSC studies with the analysis of their host-galaxy properties is needed to clarify whether mergers are required to form stellar structures at galaxy centers or simply represent one of several possible formation pathways. Whether mergers contribute primarily to the initial formation of a first-seed nucleus or instead to the subsequent growth of an already existing structure remains an open question. 

On the other hand, if the merger leads to a merger of MBHs, this may destroy the NSC. This is expected to explain the absence of NSC in massive elliptical galaxies, but needs additional studies to confirm this scenario. 
Regarding the galactic environment, while the fraction of nucleated dwarf galaxies in dense environments such as galaxy clusters is high, these galaxies are quenched right after the infall into the cluster. An interesting question is how these two facts connect. Could the formation of the NSC represent the final episode of star formation before quenching?
And why do some dwarf galaxies remain non-nucleated? Did they never form a NSC or was this destroyed by an unknown mechanism? 

During our workshop, the question of whether a bar may affect the star cluster inspiral, and thus the NSC formation through this mechanism, emerged several times, and requires a systematic investigation of the coexistence statistics of bars and NSCs. Related to inner bars, their role in funneling gas to the center and thus either feeding the central MBH or triggering in-situ star formation responsible for NSC formation or growth, is still uncharted territory and needs to be investigated further. 

Systematic studies of NSCs are essential to answer all these questions, but they are clearly difficult with current facilities since they resolve them only in very nearby galaxies (see example of spatially resolved kinematics of NSCs in \citealt{Pinna2021}). 
The next generation of 30–40 m telescopes is expected to transform the field in this sense, enabling the resolution of these small objects in a large number of galaxies, and to spatially resolve their internal properties in a sample of hundreds of galaxies within 20~Mpc. There are already proposals in this direction, both simulating future spatially resolved observations of NSCs with the ELT \citep{Nguyen2025} and presenting a science case supporting a potential extremely large telescope in the Northern Hemisphere, focused on spatially resolved properties of NSCs in over 300 galaxies within 20~Mpc \citep{Pinna2025}. 

These new studies will complement current JWST photometric and morphological studies of NSCs such as the one by \citet{Hoyer2023}, and will provide key constraints about their morphology, internal dynamics and formation history. 
In the very near future, a special focus should be dedicated to exploiting further JWST data. More generally, JWST provides complementary IR photometry for embedded dusty systems. MOONS@VLT will help constrain the chemistry at NSC spatial scales, while BlueMUSE will provide more accurate constraints for young ages. 

On a longer time scale, the JWST -- ELT synergy will allow to resolve stellar populations in extragalactic globular clusters and NSCs, 
enabling precise studies of the formation of multiple stellar populations and internal dynamics in different environments. 
In particular, integral-field spectrographs mounted on 30-40 m telescopes will allow to resolve NSCs in large samples of galaxies and to spatially resolve their internal kinematics and stellar populations in statistically significant samples of not so nearby galaxies (science cases have been proposed on these lines, see \citealp{Pinna2025}).  
Additionally, extremely large telescopes will allow to determine the bar fraction and NDR occupation fraction for galaxies at high redshift. 
For the Milky Way center, ANDES@ELT will allow accurate determination of metallicities and chemical abundances.

From the simulation perspective (see also Sect.~\ref{sub:nsc_simul}), most cosmological simulations do not resolve NSC scales yet, and this is clearly one way to go. 
In parallel to observations, new hydrodynamic simulations should model NSC formation without pre-existing black-hole or NSC seeds, including gas accretion, feedback, and cluster inspiral to reproduce observed stellar diversity and allow NSC formation to emerge self-consistently, without imposing ad hoc BH or NSC seeds. Such simulations should follow gas accretion, star formation, feedback, and cluster inspiral at sufficiently high spatial resolution to capture the formation of the first nuclear structures.
 
In the meantime, the EUCLID mission may provide important constraints on the occupation fraction of NSCs across different environments. The connection between the so-called ''little red dots''  \citep{Kocevski2023, Matthee2024, Maiolino2024} and NSCs, at the moment extensively investigated with JWST, will also help us understand the formation of the first seeds of nuclear structures and will prepare us for upcoming ELT studies in this direction. 
Despite the challenge, some other existing instruments might contribute important insights into NSC formation. MAVIS@VLT (adaptive-optics assisted) can be used for more detailed studies of the center of potential stripped NSCs to obtain a more complete chemical pattern for them (see the example of Omega Cen, Sect.~\ref{subsec:stripped_galaxy_nuclei}. In addition, WEAVE and 4MOST could advance further understanding in multiple populations, which could reveal the role of the globular-cluster inspiral scenario in NSC formation. 
MAVIS and 4MOST will contribute to expanding chemical and dynamical mapping of nuclear systems and massive clusters. 
Finally, ALMA could be useful to probe the cold gas within nearby NSCs to provide constraints on the in-situ scenario of NSC formation. 

Currently, detailed studies of circumnuclear rings and young massive star cluster formation are limited to the nearest galaxies where the $\sim100$~pc scale rings and $\sim1$~pc scale YMCs can be spatially resolved with facilities such as ALMA and JWST. The next generation of facilities, such as the SKA, ngVLA, and 30m-class telescopes, will enable resolved studies of circumnuclear rings and YMCs out to the Virgo Cluster and beyond.

\textbf{BHs}. There are still open questions about the connection between NSCs and stripped nuclei and central MBHs or IMBHs. 
For IMBHs, there are not many observational constraints yet. This is due to observational challenges related to spatial resolution, in detecting them and measuring their mass from X-ray observations or from dynamical models. 
In the case of $\omega${\,}Centauri, the population of fast-moving stars—key tracers of the IMBH—is still poorly constrained. Whether their spatial distribution is isotropic remains unclear because of the limited number of currently known fast-moving stars. These stars are also relatively faint, raising the question of whether an overabundance of low-luminosity stars is expected in the vicinity of the IMBH. 

Ongoing and future observational programs aim to address these limitations. 
These include JWST and ALMA observations targeting potential accretion signatures, JWST/NIRSpec measurements of line-of-sight velocities for fast-moving stars, and JWST/NIRCam imaging (e.g. GTO program 8322) to achieve improved spatial resolution. 
SKA will provide the radio continuum, characterization of outflows and gas accretion and inflow from large-scale structures to galaxies. 
In addition, LSST will help with the faintest objects and contribute in terms of BH demographics, while identifying multi-messenger sources as future targets for gravitational waves. 
Potential future research lines to unveil the past of the progenitors of stripped nuclei would be multiple stellar populations and the characterization of MBHs and their scaling relations. Comparing multiple-population chemistry (e.g. N-rich with or without Al enrichment) would allow to distinguish in-situ versus accreted origins.
The LISA (Laser Interferometer Space Antenna) mission is expected to provide a major step forward in the field of gravitational waves, extending the frequency detection range. This will allow us to constrain the formation of IMBHs from mergers of stellar-mass BHs, while tidal disruption events could help to observe quiescent IMBHs. 

The community would like to see simulations and modeling focusing on seeding mechanisms such as population-III-star remnants, direct collapse, or mergers in dense stellar systems. 
In the future, we should focus on the role of IMBHs in galaxy evolution, as this is key to understanding galaxies. Interesting research lines would involve, e.g., 
potential AGN feedback produced by growing IMBHs (e.g., in dwarf galaxies, not well studied since these galaxies are often too metal-poor for CO detections), the impact of runaway collisions on gas and stars. 

\textbf{BH - NSC connection}. Several open questions remain regarding the connection between MBHs, NSCs,
and dense stellar systems. Cases such as Henize~2-10 suggest that the central MBH and the surrounding NSC can follow partly decoupled formation histories, raising the question of how often MBH formation precedes the bulk of NSC
assembly \citep{Arca-Sedda2015}.
Conversely, the apparent absence of a central MBH in M~33 leaves open whether
a seed never formed, whether one formed but was not retained, or whether it
formed but subsequently failed to grow. 
A further question is whether the most informative scaling relations between MBHs and their host nuclei extend beyond the familiar $M_{\mathrm{BH}}$--$M_{\mathrm{NSC}}$ and $M_{\mathrm{BH}}$--$M_{\star}$ relations: the ratio
$r_{\mathrm{SOI}}/r_{\mathrm{h}}$ between the BH's sphere of influence
and the NSC half-mass radius, for example, may offer a more direct measure of the dynamical importance of the MBH within its host cluster.

The existence of NSCs at high redshift remains speculative, since they have not been observed and if they exist, they would certainly be very challenging to observe. 
On the other hand, high-redshift observations have shown more massive BHs than what was predicted by scaling relations obtained for low redshift, challenging previous ideas regarding the formation of MBH seeds. Whether this is related to an observational bias in high redshift BHs, or if this is true, is being investigated with current facilities. 
However, only with the ELT will we be able to systematically study the coexistence of NSCs and central BHs in dwarf galaxies, expanding the statistics with spatially resolved kinematics around the BH. NSC ages when these coexist with a central BH can also provide insights into whether the former can act as seeds for the seeding of the latter. 
MICADO@ELT will address the problem of crowded fields, for instance, in the Milky Way center, in external NSCs in which we will be able to resolve stars, and in the center of potentially stripped nuclei in the Local Group. 
HARMONI's spatial resolution will allow for the dynamical detection of BHs. 
Pushing further the use of JWST will help advance our knowledge of BH seeding. Some proposals are the imaging of the Galactic centre at larger scales (not just the very central region), tracing nuclear structures and their connection with AGNs at higher redshift, observations of larger core radius galaxies to understand how the MBHs affect core regions, further investigation of H$_2$ lines (tracing warm molecular gas), to map gas motion in central regions of galaxies.

Future gravitational-wave observations will help address these questions.
\textit{LISA} and \textit{TianQin} will probe MBH binaries at
wider separations and earlier inspiral stages than currently accessible,
allowing merger populations to be linked to their galactic and nuclear
environments and, potentially, electromagnetic counterparts to be identified
by localizing candidates in advance of coalescence and monitoring their hosts
before, during, and after merger \citep[][and references therein]{LISA2023,TianQin2025}.
Detectors operating in the decihertz regime could further open new ways to
study tidal disruption events and mergers involving IMBHs, complementing the
frequency bands accessible to \textit{LISA} and to ground-based instruments
\citep{2021ExA....51.1427A}. Such observations can place direct constraints on
MBH seeding through the merger rates, mass ratios, spins, and redshift
distributions of IMBH and low-mass MBH binaries, and help distinguish between
light-seed, heavy-seed, and stellar-dynamical formation channels. They may
also clarify whether a significant fraction of MBH mergers
originate in NSCs or other dense nuclear environments, further strengthening the link between compact stellar nuclei and MBH assembly.


\textbf{AGNs}. 
The coexistence of AGNs and nuclear stellar structures can provide important pieces of information about mechanisms that fuel gas to the galaxy center, about the balance between AGN negative and positive feedback, and other large-scale processes that might affect the center of galaxies. 
However, there are technical challenges to separating the AGN and stellar contributions. Achieving such a separation requires very high spatial resolution and will greatly benefit from the capabilities of the ELT. 

\textbf{The Milky Way center. }
Interstellar extinction remains challenging for observational studies of the stellar structures at the Milky Way center. However, JWST deep observations will extend down to 1.3 microns and allow further progress in the study of the Galactic NSC and NSD.
Concerning the NSC, detailed abundances of the NSC and NSD of the MW will provide a more complete picture of the chemical evolution and mass assembly of these structures (see also Sect.~\ref{subsec:cmz}).

\section{Conclusions}

During our workshop ''Galactic Centers as tracers of galaxy evolution'', we addressed key questions regarding stellar and gaseous structures located at the center of galaxies and processes linking these central regions to galaxy evolution on global scales.
The main highlights from the workshop are:
\begin{itemize}
    \item Current evidence suggests that NDRs share a common origin. Nuclear rings are just different stages of the inside-out formation of nuclear disks. NDRs can be used to date bars. 
    \item Circumnuclear rings lie inside the Inner Lindblad Resonance. Their gas is mainly molecular and they often host the most extreme star formation in a galaxy, leading to the formation of young massive star clusters and driving strong stellar feedback sometimes with outflows that reach the galaxy scale. 
    \item Some inner bars are formed out of stars belonging to nuclear stellar disks, analogous to the dynamical formation of larger-scale stellar bars. Inner bars are long-lived and they can buckle and develop boxy/peanut structures. They can funnel gas to the very central regions of galaxies, further inwards than allowed by a regular-sized single bar. 
    \item The formation of NSCs in dwarf galaxies is sensitive to the galaxy environment, as shown by their occupation fraction, and a connection between the observed occupation fraction and the halo mass of a galaxy cluster has been found. Gas-rich mergers can lead to gas inflows to the center and in-situ star formation, leading to NSC formation. Hierarchical galaxy mass assembly can also form NSCs through star-cluster mergers. Different types of simulations support both in-situ formation and star-cluster inspiral as formation mechanisms for NSCs, although the dominant formation channel in different galaxy types remains unclear. 
    \item Some nearby spiral galaxies apparently lack a central MBH (or this cannot be detected with current instruments). The potential absence of a MBH can be due to inefficient seeding or growth, or displacement or ejection of a previous MBH. MBH seeding can happen through three channels: remnants of Population III stars producing light seeds, direct collapse producing heavy seeds (explaining high-redshift observations), and processes fostered by dense stellar environments, producing IMBHs (e.g., runaway stellar collisions, stellar-mass BH mergers).
    \item NSCs seem to play a role in BH seeding and growth, providing a dense environment favorable for seeding, and funneling gas to the central parsecs, helping BH growth. 
    \item While stellar feedback has been traditionally considered as responsible for the regulation of star formation in low-mass galaxies, it has been recently shown that AGN feedback can be more effective than stellar feedback at clearing gas from low-mass galaxies, and that a subset of this galaxy population hosts accreting MBHs. Some simulations have revealed that AGN feedback in dwarfs suppresses star formation primarily through episodic heating, rather than gas removal, often leading to dual-mode cycles: combined heating and outflows, alternating with shocks and compression and subsequent star formation. More generally, the link between AGN activity and circumnuclear star formation has suggested that BH feeding, starbursts, and AGN feedback are just different steps of the same evolutionary sequence. 
    \item While the SMBH at the center of the Milky Way, Sgr A*, currently has a low luminosity, aside from some variability, it experienced significant activity only a few hundred years ago.
    \item The CMZ gas ring grows slowly in radius, building the Milky Way NSD inside out. Intense star formation or magnetic fields might be responsible for inflows from the further CMZ to the very center of our Galaxy.
    \item Stripped nuclei can provide important insight into NSC formation, BH seeding and the evolution of dwarf galaxies. An IMBH has been recently detected at the center of $\omega$~Cen. 
\end{itemize}

The following questions, on the other hand, remain unanswered:
\begin{itemize}
    \item The connection between NSCs, NDRs and bars is unclear. They do not share the same scaling relations. 
    \item The dominant formation pathways of NSCs in late-type galaxies remain poorly constrained.
    \item There is no general agreement on the occupation fraction of MBHs in galactic nuclei, particularly unclear at the low-mass galaxy end and in spiral galaxies of all masses. The fraction of central BHs in low-mass galaxies remains essential to understanding the origin of MBH seeds. 
    \item Demographics and origin of IMBHs are not constrained, as well as the low-mass end of the MBH population, because they are currently detected only when observations have sufficient spatial resolution and sensitivity.
\end{itemize}
The following research directions were identified as priorities for addressing these open questions.

\begin{itemize}
    \item Systematic observational studies of NSCs, NDRs, and nuclear bars, in a large sample of galaxies covering the low-mass end and the latest-type spirals. These systematic studies will include spatially resolved spectroscopic properties of these small structures, thanks to integral-field spectrographs mounted on the new-generation 30-40-m telescopes. 
    \item Cosmological simulations capable of resolving down to NSC spatial scales.
    \item Cosmological simulations without relying on ad hoc seeding of nuclear structures and BHs.
    \item Extension of resolved studies of circumnuclear rings and YMCs out to beyond the Virgo cluster, thanks to the next-generation radio facilities. 
    \item Targeting potential BH accretion signatures and constraining velocities of fast-moving stars to improve BH demographics. 
    \item Identifying potential multi-messenger sources and characterizing them with future gravitational-wave observatories.
    \item Probing MBH binaries and characterizing BH mergers through gravitational-wave observations. 
    \item Tracing properties of stripped-nuclei progenitors through multiple stellar populations and BHs. 
    \item Achieving detailed chemical abundances of stars in different components of the Milky Way center. 
    \item Advancing our understanding of the AGN–stellar nucleus connection by disentangling their respective contributions through the increased spatial resolution provided by the next generation of extremely large telescopes.

\end{itemize}


\textit{The remaining author affiliations are:}

{
$^{3}$ LIRA, Observatoire de Paris, Universit{\'{e}} PSL, CNRS, Sorbonne Universit{\'{e}}, Universit{\'{e}} Paris Cit{\'{e}}, CY Cergy Paris Universit{\'{e}}, 5 place Jules Janssen, 92195 Meudon, France
 
$^{4}$ Department of Astrophysics, University of Vienna, T\"urkenschanzstraße 17, 1180 Vienna, Austria

$^{5}$ Dipartimento di Fisica e Astronomia “Galileo Galilei”, Università di Padova, Vicolo dell’Osservatorio 3, 35122 Padova, Italy 

$^{6}$ Dipartimento di Tecnica e Gestione dei Sistemi Industriali, Università di Padova, Stradella S. Nicola 3, I-36100 Vicenza, Italy 

$^{7}$ INAF – Osservatorio Astronomico di Padova, Vicolo dell’Osservatorio 5, Padova, I-35122, Italy 

$^{8}$
Nicolaus Copernicus Astronomical Center, Polish Academy of Sciences, Warsaw, Poland

$^{9}$ Space Telescope Science Institute, 3700 San Martin Drive, Baltimore, MD 21218, USA

$^{10}$
Instituto de Astrofísica de Andalucía (CSIC), Glorieta de la Astronomía S/N, 18008 Granada, Spain

$^{11}$ European Southern Observatory, Alonso de Córdova 3107, Vitacura, Región Metropolitana, Chile

$^{12}$ Donostia International Physics Centre, Paseo Manuel de Lardizabal 4, 20018 Donostia-San Sebastian, Spain

$^{13}$ IKERBASQUE, Basque Foundation for Science, E-48013, Bilbao, Spain

$^{14}$
Leibniz-Institute for Astrophysics Potsdam (AIP), An der Sternwarte 16, D-14482 Potsdam, Germany

$^{15}$
Division of Astrophysics, Department of Physics, Lund University, Box 118, SE-22100 Lund, Sweden
}

\begin{acknowledgements}
The authors would like to thank the Lorentz Center for hosting the workshop ''Galactic centers as tracers of galaxy evolution'', especially Maria Krebbers, who made sure that everything worked well. 
F.P. acknowledges support from the Horizon Europe research and innovation programme under the Maria Sklodowska-Curie grant “TraNSLate” No 101108180. A.F.K. acknowledges funding from the Austrian Science Fund (FWF) [grant DOI 10.55776/ESP542]. AA acknowledges that this research was funded in part by National Science Centre (NCN), Poland, grant No. 2024/55/D/ST9/02585. For the purpose of Open Access, the author has applied a CC BY public copyright licence to any Author Accepted Manuscript (AAM) version arising from this submission. RS acknowledges support by the Severo Ochoa grant CEX2021-001131-S, funded by MCIN/AEI/10.13039/501100011033, and from grant PID2022-136640NB-C21 funded by MCIN/AEI 10.13039/501100011033 and by the European Union.
\end{acknowledgements}

\bibliographystyle{aa}
\bibliography{bibliography}


\end{document}